\documentstyle[12pt]{article}

\newcommand{\nc}{\newcommand}

\nc{\dbar}{\bar{\partial}}

\catcode`\@=11

\@addtoreset{equation}{section}
\def\theequation{\thesection\arabic{equation}}

\def\@normalsize{\@setsize\normalsize{15pt}\xiipt\@xiipt
\abovedisplayskip 14pt plus3pt minus3pt%
\belowdisplayskip \abovedisplayskip
\abovedisplayshortskip  \z@ plus3pt%
\belowdisplayshortskip  7pt plus3.5pt minus0pt}
\def\small{\@setsize\small{13.6pt}\xipt\@xipt
\abovedisplayskip 13pt plus3pt minus3pt%
\belowdisplayskip \abovedisplayskip
\abovedisplayshortskip  \z@ plus3pt%
\belowdisplayshortskip  7pt plus3.5pt minus0pt
\def\@listi{\parsep 4.5pt plus 2pt minus 1pt
            \itemsep \parsep
            \topsep 9pt plus 3pt minus 3pt}}

\def\underline#1{\relax\ifmmode\@@underline#1\else
        $\@@underline{\hbox{#1}}$\relax\fi}
\@twosidetrue
\relax

\catcode`@=12

\catcode`\@=11

\def\section{\@startsection{section}{1}{\z@}{3.5ex plus 1ex minus
   .2ex}{2.3ex plus .2ex}{\large\bf}}

\def\ps@headings{\def\@oddfoot{}\def\@evenfoot{}
\def\@oddhead{\hbox{}\hfill
        \makebox[.5\textwidth]{\raggedright\ignorespaces --\thepage{}--
        \hfill }}
\def\@evenhead{\@oddhead}
\def\subsectionmark##1{\markboth{##1}{}}
}

\ps@headings

\catcode`\@=12

\relax

\def\figcap{\section*{Figure Captions\markboth
        {FIGURECAPTIONS}{FIGURECAPTIONS}}\list
        {Fig. \arabic{enumi}:\hfill}{\settowidth\labelwidth{Fig. 999:}
        \leftmargin\labelwidth
        \advance\leftmargin\labelsep\usecounter{enumi}}}
 \relax
\def\tablecap{\section*{Table Captions\markboth
        {TABLECAPTIONS}{TABLECAPTIONS}}\list
        {Table \arabic{enumi}:\hfill}{\settowidth\labelwidth{Table 999:}
        \leftmargin\labelwidth
        \advance\leftmargin\labelsep\usecounter{enumi}}}
 \relax
\def\reflist{\section*{References\markboth
        {REFLIST}{REFLIST}}\list
        {[\arabic{enumi}]\hfill}{\settowidth\labelwidth{[999]}
        \leftmargin\labelwidth
        \advance\leftmargin\labelsep\usecounter{enumi}}}
 \relax

\catcode`\@=11

\def\ps@headings{\def\@oddfoot{}\def\@evenfoot{}
\def\@oddhead{\hbox{}\hfill
        \makebox[.5\textwidth]{\raggedright\ignorespaces --\thepage{}--
        \hfill }}
\def\@evenhead{\@oddhead}
\def\subsectionmark##1{\markboth{##1}{}}
}

\ps@headings

\relax

\def\firstpage#1#2#3#4#5#6{
\begin{document}

\begin{titlepage}
\nopagebreak
\title{\begin{flushright}
        \vspace*{-1.2in}
        {\normalsize TUM--HEP--268/97 #2
}\\[-5mm]
        {\normalsize hep-th/9703047}\\[-5mm]
{\normalsize March 1997}\\[.5cm]
\end{flushright}
\vspace{2cm}
{\large \bf #3}}
\author{\large #4 \\ #5}
\maketitle
\vskip -7mm
\nopagebreak
\begin{abstract}
{\noindent #6}
\end{abstract}
\vfill
\thispagestyle{empty}
\end{titlepage}}
\newcommand{\dal}{\raisebox{0.085cm}
{\fbox{\rule{0cm}{0.07cm}\,}}}
\newcommand{\bb}{\begin{eqnarray}}
\newcommand{\ee}{\end{eqnarray}}
\newcommand{\p}{\partial}
\newcommand{\bp}{{\bar \p}}
\newcommand{\bR}{{\bf R}}
\newcommand{\bC}{{\bf C}}
\newcommand{\bZ}{{\bf Z}}
\newcommand{\bS}{{\bar S}}
\newcommand{\bT}{{\bar T}}
\newcommand{\bU}{{\bar U}}
\newcommand{\bA}{{\bar A}}
\newcommand{\bh}{{\bar h}}
\newcommand{\bu}{{\bf{u}}}
\newcommand{\bv}{{\bf{v}}}
\newcommand{\D}{{\cal D}}
\newcommand{\s}{\sigma}
\newcommand{\Sg}{\Sigma}
\newcommand{\ket}[1]{|#1 \rangle}
\newcommand{\bra}[1]{\langle #1|}
\newcommand{\non}{\nonumber}
\newcommand{\ph}{\varphi}
\newcommand{\la}{\lambda}
\newcommand{\ga}{\gamma}
\newcommand{\ka}{\kappa}
\newcommand{\m}{\mu}
\newcommand{\n}{\nu}
\newcommand{\dX}{\dot{X}}
\newcommand{\th}{\vartheta}
\newcommand{\Lie}[1]{{\cal L}_{#1}}
\newcommand{\eps}{\epsilon}
\newcommand{\bz}{\bar{z}}
\newcommand{\bX}{\bar{X}}
\newcommand{\om}{\omega}
\newcommand{\Om}{\Omega}
\newcommand{\we}{\wedge}
\newcommand{\La}{\Lambda}
\newcommand{\bOm}{{\bar \Omega}}
\newcommand{\CA}{{\cal A}}
\newcommand{\CF}{{\cal F}}
\newcommand{\CbF}{\bar{\CF}}
\newcommand{\CAM}{\CA^{(M)}}
\newcommand{\CAS}{\CA^{(\Sg)}}
\newcommand{\CFS}{\CF^{(\Sg)}}
\newcommand{\I}{{\cal I}}
\newcommand{\al}{\alpha}
\newcommand{\be}{\beta}
\newcommand{\cm}{Commun.\ Math.\ Phys.~}
\newcommand{\pr}{Phys.\ Rev.\ D~}
\newcommand{\pl}{Phys.\ Lett.\ B~}
\newcommand{\ibar}{\bar{\imath}}
\newcommand{\jbar}{\bar{\jmath}}
\newcommand{\np}{Nucl.\ Phys.\ B~}
\newcommand{\e}{{\rm e}}
\newcommand{\gsi}{\,\raisebox{-0.13cm}{$\stackrel{\textstyle
>}{\textstyle\sim}$}\,}
\newcommand{\lsi}{\,\raisebox{-0.13cm}{$\stackrel{\textstyle
<}{\textstyle\sim}$}\,}
\date{}
\firstpage{97/XX}{}
{\Large\sc  On the Ground State of the Supersymmetric Five--Brane
}
{ Alexandros Kehagias
}
{ \normalsize
 {\it Physik Department}\\
\normalsize {\it Technische Universit\"at M\"unchen} \\
\normalsize {\it D-85748 Garching, Germany}\\
\normalsize {\it E-mail address: kehagias@physik.tu-muenchen.de}}
{\normalsize
We examine   if there exists a zero--energy supersymmetric
ground state for the fundamental five--brane.
Looking for  an $SO(6)\!\times\!SO(2)$--invariant ground state,
we construct, in the light--cone gauge,  perturbatively a Nicolai map
up to third order
in the inverse five--brane tension.
We show that the Nicolai map equilibrates and the five--brane has  a
zero--energy normalizable supersymmetric vacuum state.
For the other p--branes, we argue that only the
three--brane has a  zero--energy ground state.
}

\newpage

\section{Introduction}
\pagestyle{plain}

The existence of extended objects, p--branes, in the string
spectrum promises that interesting information about the
non-perturbative structure of string theory can be obtained. These
p--branes configurations appear as solitons in the low--energy string
field theory and  they are necessary in establishing
the various string dualities.
There exists extensive literature dealing with their
properties and their dynamics \cite{DD}--\cite{Duff}.

On the other hand, solitonic p--branes
are quit different in many aspects from fundamental ones \cite{DD}--\cite{BST}.
For example, solitonic branes have internal structure, lost in the long
wave--length limit, while by definition there is no structure for
fundamental p--branes. The excitations of the latter are
interpreted as ordinary particles. Thus, in order for one to really
think of extended objects as being fundamental, the question of the
existence of massless states in the spectrum has to be answered.
This issue, as far as we know,  has extensively be studied only
for  the  fundamental membrane \cite{WLN}--\cite{W}.
However, the same question should
also be addressed  for the other extended objects which admit space--time
supersymmetry, namely, for the three--, four-- and five--branes.
Here we will consider explicitly the latter and in particular the neutral ones
since the heterotic five--brane action is not known.
We will also make some comments on the other p--branes.

There exists a serious difference between the five--brane and
the fundamental membrane. For the latter, there exist supersymmetric  quantum
mechanical models with finite degrees of freedom
for which the Schr\"odinger equation can explicitly
be studied. These models are supersymmetric $SU(N)$ Yang--Mills
theories dimensionally reduced to $0\!+\!1$ dimensions (time).
The fundamental membrane is then recovered in the
$N\!\rightarrow\!\infty$ limit \cite{Hop}.
For this class of models, the
result is that the spectrum is continuous
starting from zero and filling the positive real axis \cite{WLN}.
Moreover, there is no normalizable zero--energy
state. This is  consistent with the proposal that the membrane is
effectively described by condensation of D0 branes of the type IIA theory
\cite{TT},\cite{SS}.
However, it should  be mentioned that there exists also the claim that
the membrane has discrete spectrum \cite{TTT} as well, as a consequence of
the finite size core of the D--particle \cite{Bachas}.

For the five--brane on the other hand \cite{D1},\cite{D2},
there is no quantum mechanical
model and so one is forced to study a system with infinite degrees of
freedom.  In this case, instead of solving functional differential equations,
we prefered to follow another way,
namely, to find a  Nicolai
map \cite{Nic} perturbatively in the inverse five--brane tension.
We determined such a
map in the case of $SO(6)\!\times\!SO(2)$--symmetric target space.
Moreover, the Nicolai map we have constructed equilibrates for large
times and thus there exists a zero--energy supersymmetric vacuum
state. This state corresponds to the
$N=1$ ten--dimensional vector multiplet \cite{BPS}.
\vspace{.2cm}

In sect. $2$ we set up the formalism and we give the
partition function for the five--brane in the light--cone gauge. Next we
discuss the Nicolai map in sect. $3$ and in sect. $4$ we find the
ground state wave function  and we make some
comments for the corresponding state of three-- and four--branes.
Finally, we summarize our results in sect. 5.

\section{The five-brane action}

Quantization  in the light--cone gauge is sometimes convenient  since
unitarity is guaranteed. A drawback is that Lorentz invariance may be lost
as a consequence of quantization and should be checked at the end. It is also
possible that this gauge is the only one in which a Hamiltonian formulation
of a  theory can be performed as for example the string theory. Here we will
study the supersymmetric five--brane in the light--cone gauge which is
described by the  Lagrangian \cite{BSTT}
\bb
{\cal{L}}&=&\frac{T_5}{2}\left(\D X^I\D X^I
-det\partial_a
X^I\partial_bX^I+i\bar{S}\D S \right.\nonumber \\
&&\left.+\frac{i}{4!}
\varepsilon^{abcde}\partial_aX^I\p_bX^J\p_cX^K\p_dX^L\bar{S}
\gamma_{IJKL}\partial_eS\right)\, . \label{5brane}
\ee
$T_5$ is the five--brane tension and we will assume, if it is not explicitly
indicated that $T_5=1$.
The covariant derivative $\D$ is given by
 $$\D=\partial/\partial t+u^a\partial/\partial\sigma^a ~~~
(a,b=1,\cdots,5)\, ,$$
where $\sigma^a$ are coordinates on the brane and $u^a$ is a
divergence free
\bb
\partial_au^a=0\, , \label{u}
\ee
vector field.  In
the light--cone gauge only $d-2$ of the original $d$  fields
remain and
since a supersymmetric five--brane may live only in $d=10$,  there exists
eight bosonic fields
$X^I,~ (I=1,\cdots,8)$.
However, as we will see below, due to gauge symmetries
only the transverse excitations of the brane remain which  represent the
physical degrees of freedom.
The fermion $S$ is a real $SO(8)$ spinor which
 we will assume to be the
$\bf{8_c}$  and $\bar{S}=S^T$.

The action for the five--brane is invariant under the
 supersymmetry transformations
\bb
\delta X^I&=&2i\bar{\epsilon}\gamma^IS\, ,\nonumber \\
\delta S&=&-2\D X^I\gamma_I\epsilon+\frac{2}{5!}
\varepsilon^{abcde}\partial_aX^I\partial_bX^J
\p_cX^K\p_dX^L\p_eX^M\gamma_{IJKLM}\epsilon\, ,\nonumber \\
\delta u^a&=&-\frac{i}{3}\bar{\epsilon}\gamma_{IJK}
\varepsilon^{abcde}\p_bX^I\p_cX^J\p_dX^K
\p_eS \, , \label{supersymmetry5}
\ee
where $\gamma^I$ are $SO(8)$ $\gamma$--matrices and $\epsilon^{abcde}$ is the
totally antisymmetric symbol in five dimensions. It is also invariant under
reparametrizations $\sigma^a\rightarrow \sigma^a+\epsilon^a$
 and  the fields in (\ref{5brane}) transform as
\bb
\delta X^I&=&\varepsilon^a\partial_aX^I\, ,\nonumber \\
\delta S&=&\varepsilon^a\partial_aS\, ,\nonumber \\
\delta u^a&=&-\frac{d\varepsilon^a}{dt}+
\varepsilon^b\partial_bu^a-u^b\partial_b\varepsilon^a \, . \label{gauge}
\ee
The vector field $\varepsilon^a$  generates
diffeomorphisms on the five--brane and due to the constraint eq.(\ref{u}),
 $\varepsilon^a$ is also divergence free
$$\partial_a\varepsilon^a=0.$$
Hence, $\varepsilon^a$
generates, in fact,  volume preserving
diffeomorphisms.
We will gauge fix the  reparametrization invariance  by choosing the gauge
\bb
u^a=0\, . \label{gfixing}
\ee
 Then ghosts $c^a$ and anti--ghosts $\bar{c}^a$
are also divergence free, i.e. they satisfy  the condition
$$
\partial_ac^a=\partial_a\bar{c}^a=0,
$$
which may implemented in the
action by Lagrange multipliers $(\lambda,\bar{\lambda})$. The
Faddeev--Popov determinant for the gauge fixing eq.(\ref{gfixing}) is
$$\Delta_{FP}=det\left(\frac{d}{dt}\delta_{ab}\delta(\sigma\!-\!\sigma')
\delta(t\!-\!t')\right),$$
and the ghost action is then
\bb
I_{gh}=\int dtd^5\sigma \left( ic^a\frac{d\bar{c}^a}{dt}+i\bar{\lambda}
\partial_ac^a+{\tt {h.c.}}
\right)\, . \label{gh}
\ee

Let us now introduce the Nambu bracket \cite{Nambu}
\bb
\{X^{I_1},\cdots,X^{I_5}\}=\varepsilon^{{a_1}\cdots{a_5}}\partial_{a_1}X^{I_1}
\cdots\partial_{a_5}X^{{I_5}} \, , \label{Nambu}
\ee
which is skew--symmetric, satisfies the Leibniz rule and the fundamental
identity \cite{ta}
\bb
\{\{X^{{I_1}},\cdots,X^{{I_5}}\},X^{{I_6}},\cdots,X^{{I_9}}\}+
\{X^{{I_5}},\{X^{{I_1}},\cdots,X^{{I_4}}
X^{{I_6}}\},X^{{I_7}},X^{{I_8}},X^{{I_9}}\}
\nonumber \\
+\cdots+\{X^{{I_5}},\cdots,X^{{I_8}},\{X^{{I_1}},\cdots,X^{{I_4}},X^{{I_9}}\}\}
=\{X^{{I_1}},\cdots,X^{{I_4}},\{X^{{I_5}},\cdots,X^{{I_9}}\}\} \, , \nonumber
\ee
which is a generalization of the Jacobi identity.
The determinant in the
Lagrangian (\ref{5brane}) may be expressed in terms of the Nambu bracket as
\bb
det(\partial_aX^I\partial_bX^I)=\frac{1}{5!}
\{X^{I_1},\cdots,X^{I_5}\}^2\, .
\label{det}
\ee
The five--brane action turns out then to be
\bb
I\!\!=\!\frac{1}{2}\!\int dt d^5\sigma\!\left((\D X^I)^2
\!-\!\frac{1}{5!}\!\{X^{I_1},\cdots,X^{I_5}\}^2\!+
\!i\bar{S}\D S\!+\!\frac{i}{4!}
\bar{S}
\gamma_{IJKL}\{X^I,X^J,X^K,X^L,S\}\right)\, , \label{5brane1}
\ee
and the equations of motions  as follow from (\ref{5brane1}) are
\bb
\frac{d^2X^I}{dt^2}-
\frac{1}{4!}\{\{X^I,X^{I_1},\cdots,X^{I_4}\},X^{I_1},\cdots,
X^{I_4}\}-\frac{i}{3!}\{\bar{S}{\gamma^I}_{JKL},X^J,X^K,X^L,S\}=0,\\
\frac{dS}{dt}+\frac{1}{4!}\gamma_{IJKL}\{X^I,X^J,X^K,X^L,S\}=0 \, . \label{eq}
\ee

In order to discuss quantum aspects of the fundamental five--brane,
we will consider the partition function of the theory  which we write as
\bb
Z=\int d\mu e^{-I_E-I_{gh}} \, .
\ee
The measure $d\mu$ is
$$ d\mu=[dX^I][d\bar{S}][dS][d\bar{c}][dc][d\bar{\lambda}][d\lambda]$$
and
\bb
I_E\!=\!\frac{1}{2}\!\int d\tau d^5\sigma\!\left((\frac{dX^I}{d\tau})^2
\!+\!\frac{1}{5!}\!\{X^{I_1},\cdots,X^{I_5}\}^2\!+
\!\bar{S}\frac{dS}{d\tau}\!-\!\frac{i}{4!}
\bar{S}
\gamma_{IJKL}\{X^I,X^J,X^K,X^L,S\}\right)\!\!, \label{5braneE}
\ee
is the gauge fixed Euclidean five--brane action after a
Wick rotation of (\ref{5brane1}).
Integrating out the ghosts  and the Lagrange multiplier  we get
\bb
Z=\int [dX^I][d\bar{S}][dS] e^{-I_E}
det\left(\frac{d}{d\tau}\delta(\sigma\!-\!\sigma')
\delta(\tau\!-\!\tau')\right)^4
det(\partial_a\partial^a)\, , \label{int}
\ee
and thus, as advertised,  only four of the $X^I$ represent
the physical excitations of the brane.

 We may also integrate out  the fermions  which appear
quadratically in (\ref{5braneE}). Thus, finally, the partition function may be
expressed as an integral over bosonic fields only as
\bb
Z=\int [dX^i]e^{-I_E[X]}
det\left(\frac{d}{d\tau}\delta(\sigma\!-\!\sigma')
\delta(\tau\!-\!\tau')\right)^4
det_F\, ,
\label{Z}
\ee
where
\bb
det_F=
det\left[\left(\frac{d}{d\tau}\delta_{\alpha\beta}
+\frac{i}{4!}(\gamma_{IJKL})_{\alpha\beta}\{X^I,X^J,X^K,X^L,~~\}\right)
\delta(\sigma\!-\!\sigma')
\delta(\tau\!-\!\tau')\right]^{1/2}
\ee
is the fermionic determinant and
$(\alpha,\beta=1,\cdots,8)$ are spinor indices.
It should be noted
here that we assume  periodic boundary conditions for both
bosons and fermions in order supersymmetry to be respected.

Let us suppose now that it is possible
to find a transformation $X^I\rightarrow \xi^I(X)$ such that i) the Jacobian of
this transformation  cancels exactly the product of determinants in
eq.(\ref{Z}) and ii) $I_E$ is proportional to the length $|\xi|^2$. Such
transformation, known as Nicolai map, reduces the partition function into a
Gaussian integration.
It is in general a non--local and
non--polynomial transformation which, however,
can be constructed order by order in perturbation theory
\cite{Nic},\cite{st}. Exact expressions may
be obtained for topological field theories \cite{top}. In the next
section we will see that such a $\xi^I(X)$ can be found approximately up to
third order in $1/T_5^2$ in a similar way as in
the four--dimensional supersymmetric Yang--Mills theory \cite{Nic}.

\section{The Nicolai map}

The previous considerations were quite general. Here we will study a
particular case, namely, we will split $X^I$ as
$X^I=(X^i,X^7,X^8),~~(i=1\cdots,6)$ and we will assume that $(X^7,X^8)$ are
constants. This breaks the original $SO(8)$ symmetry into
$SO(6)\!\times\!SO(2)$. The spinor $S$ is split accordingly into
$\bf{4}+\bar{\bf{4}}$ and one may form the two real spinors
$\theta_1\sim\bf{4}\!+\!\bar{\bf{4}}$ and
$\theta_2\sim i(\bf{4}\!-\!\bar{\bf{4}})$.
Similarly, we will assume that $\theta_1$ is also constant.

We may consider  the  fields  $X^i(\sigma)$ as a  map
$X^i:\Sigma\rightarrow M$ from the five--brane worldvolume
$\Sigma$ which we take to be $S^6$
to the six--dimensional target space $M$ parametrized by $X^i$.
In this case we may define the winding number
(``degree") of this map as
\bb
q=\frac{1}{128\pi^6}
\int d\tau d^5\sigma \frac{dX^j}{d\tau}\partial_{a_1}X^{i_1}
\cdots\partial_{a_5}X^{{i_5}}\varepsilon^{{a_1}\cdots{a_5}}
\epsilon_{j{i_1}\cdots{i_5}},
\ee
which may also be written as
\bb
q=\frac{1}{128\pi^6}
\int d\tau d^5\sigma \frac{dX^j}{d\tau}\{X^{i_1},\cdots,X^{i_5}\}
\epsilon_{j{i_1}\cdots{i_5}}\, . \label{degree}
\ee

We introduce  now new bosonic fields $\xi^i$ defined by
\bb
\xi^i&=& \frac{dX^i}{d\tau}+\frac{1}{5!} \{X^{i_1},\cdots,X^{i_5}\}
{\epsilon^i}_{{i_1}\cdots{i_5}}\, ~~~if~~~~ q<0 \nonumber \\
\xi^i&=& \frac{dX^i}{d\tau}-\frac{1}{5!} \{X^{i_1},\cdots,X^{i_5}\}
{\epsilon^i}_{{i_1}\cdots{i_5}}\,  ~~~if~~~~ q\geq 0\, , \label{x}
\ee
where $\epsilon_{{i_1}\cdots{i_6}}$ is the six--dimensional antisymmetric
symbol. We will see below that the transformation $X^i\rightarrow \xi^i(X)$
is a Nicolai map. One may easily verify that
\bb
\xi^i\xi^i=(\frac{dX^i}{d\tau})^2+det(\partial_aX^i\partial_bX^i)
\pm\frac{2}{5!}\frac{dX^i}{d\tau}\{X^{i_1},\cdots,X^{i_5}\}
\epsilon_{i{i_1}\cdots{i_5}} \, , \label{xx}
\ee
for $q<0, q\geq 0$.
Then, the bosonic part of the action (\ref{5braneE}) is written as a
quadratic form in $\xi^i$,
\bb
I_E^{bos}=\frac{1}{2}\int \xi^i\xi^i +|Q| \, ,
\ee
where $Q=128\pi^6 q$.
As a result, the partition function (\ref{Z}) turns out to be
\bb
Z=\int [d\xi^i]e^{-\frac{1}{2}\int \xi^i\xi^i -|Q|}
|det(\frac{\delta\xi^i}{\delta X^j})|^{-1}
det\left(\frac{d}{d\tau}\delta(\sigma\!-\!\sigma')
\delta(\tau\!-\!\tau')\right)^4
det_F \, , \label{ZZ}
\ee
where
$|det(\frac{\delta\xi^i}{\delta X^j})| $ is the Jacobian of the map
$X^i\rightarrow\xi^i(X)$. We will show below that
\bb
det(\frac{\delta\xi^i}{\delta X^j})=
det\left(\frac{d}{d\tau}\delta(\sigma\!-\!\sigma')
\delta(\tau\!-\!\tau')\right)^4
det_F \, , \label{aa}
\ee
up to third order in $T_5^{-2}$ so that
\bb
{\cal{M}}=1+{\cal{O}}(T_5^{-8})\, .\label{MM}
\ee
with
\bb
{\cal{M}}= det(\frac{\delta\xi^i}{\delta X^j})^{-1}
det\left(\frac{d}{d\tau}\delta(\sigma\!-\!\sigma')
\delta(\tau\!-\!\tau')\right)^4
det_F  \, .\label{M}
\ee
Thus, the partition function will be transformed into  a Gaussian
integration over $\xi^i$'s, as required from a Nicolai map.

The Jacobian of  $X^i\rightarrow\xi^i(X)$ is
\bb
det(\frac{\delta\xi^i}{\delta X^j})=
det\left((\frac{d}{d\tau}\delta_{ij}
\pm\frac{1}{4!}\epsilon_{ijklmn}\{X^k,X^l,X^m,X^n,~~\})
\delta(\sigma\!-\!\sigma')\delta(\tau\!-\!\tau')\right)\, ,
\ee
which we may write as
\bb
det(\frac{\delta\xi^i}{\delta X^j})=det\left(\frac{d}{d\tau}
\delta(\sigma\!-\!\sigma')\delta(\tau\!-\!\tau')\right)^6
det({\bf{1}}\pm A)\, , \label{IA}
\ee
where $A$ is the antisymmetric matrix
\bb
A_{ij}=\frac{1}{4!}\epsilon_{ijklmn}\p_\tau^{-1}\{X^k,X^l,X^m,X^n,~~\}
\delta(\sigma\!-\!\sigma')\delta(\tau\!-\!\tau') \, ,
\ee
and ${\bf{1}}_{ij}=\delta_{ij}$.
Expanding the determinant $det({\bf{1}}\pm A)$
in the right hand side of eq.(\ref{IA}) we get
\bb
det(\frac{\delta\xi^i}{\delta X^j})=det\left(\frac{d}{d\tau}
\delta(\sigma\!-\!\sigma')\delta(\tau\!-\!\tau')\right)^6
\!\left(1\!-\!\frac{1}{2}Tr(A^2)\!+\!\frac{1}{4!}\left[TrA^4
\!-\!(TrA^2)^2\right]\!+\cdots
\right). \label{exp}
\ee
Similarly, the fermionic determinant is written as
\bb
det_F=  det\left(\frac{d}{d\tau}
\delta(\sigma\!-\!\sigma')\delta(\tau\!-\!\tau')\right)^2 det(I+\Gamma)^{1/2}
\, , \label{df}
\ee
where $I_{\bar{\alpha}\bar{\beta}}=\delta_{\bar{\alpha}\bar{\beta}}$,
$(\bar{\alpha}\bar{\beta}=1,\cdots,4)$ and
\bb
\Gamma
=\frac{i}{4!}
\bar{\gamma}_{ijkl}
\p_\tau^{-1}
\{X^i,X^j,X^k,X^l,~~\}\delta(\sigma\!-\!\sigma')
\delta(\tau\!-\!\tau') \, , \label{g}
\ee
with  $\bar{\gamma}^i$ the $SO(6)$ $\gamma$--matrices.
Expanding the determinant $det(I+\Gamma)$ in eq.(\ref{df}) we get
\bb
det_F\!=\! det\left(\frac{d}{d\tau}
\delta(\sigma\!-\!\sigma')\delta(\tau\!-\!\tau')\right)^2
\left(\!1\!-\!\frac{1}{4}Tr(\Gamma^2)\!+\!\frac{1}{48}\left[Tr\Gamma^4\!-\!
\!\frac{1}{2}(Tr\Gamma^2)^2\right]\!+\cdots\!
\right)\, . \label{expG}
\ee
It is a straightforward matter to verify that
\bb
Tr(\Gamma^2)&=&2TrA^2 \, , \nonumber \\
Tr(\Gamma^4)&=&-\frac{3}{4}\left(TrA^4-2(TrA^2)^2\right)\, .\label{GG}
\ee
Comparing then eqs.(\ref{exp},\ref{expG}) using eq.(\ref{GG}), one may verify
eq.(\ref{aa}) up to third order and thus, indeed the Jacobian of the
transformation $X^i\rightarrow\xi^i(X)$
cancels the fermionic and the Faddeev--Popov determinants.
Since the expansion of the determinants was actually an
expansion in $1/T_5^2$, eq.(\ref{MM}) follows trivially.

The transformation $X^i\rightarrow\xi^i(X)$ turns thus the partition function
into the  Gaussian integration over $\xi^i$'s
\bb
Z\sim \int[d\xi^i]e^{-\frac{1}{2}\int
\xi^i\xi^i-|Q|}sign\det(\frac{d\xi^i}{dX^i})\, ,  \label{sign}
\ee
up to cubic order.
The factor $sing\det$ above is due to the fact that it is the modulus of
$\det(\delta\xi/\delta X)$ rather than the determinant itself which appears in
eq.(\ref{ZZ}).
If we define the operator $\Delta_{ij}=\frac{\delta\xi^i}{\delta
X^j}$, we have that
\bb
\Delta_{ij}=\frac{d}{d\tau}\delta_{ij}
\pm\frac{1}{4!}\epsilon_{ijklmn}\{X^k,X^l,X^m,X^n,~~\}
\, .
\ee
One may easily verify that
\bb
sing
\det\Delta_{ij}=exp\left(i\frac{\pi}{2}
\left[\zeta_\Delta(0)-\eta_\Delta(0)\right]\right)
\, ,
\ee
where $ \zeta_\Delta(s)=\sum_n |\lambda_n|^{-s}$ and $\eta_\Delta(s)=
\sum_{n}sign(\lambda_n)|\lambda_n|^{-s}$ are the $\zeta$--function
 and the  $\eta$--invariant of $\Delta_{ij}$. Thus, finally, up to third order
\bb
Z\sim\int[d\xi^i]\exp\left(-\frac{1}{2}\int\xi^i\xi^i-|Q|+
i\frac{\pi}{2}\left[\zeta_\Delta(0)-\eta_\Delta(0)\right]\right) \,  .
\label{fin}
\ee

 What is
still missing is the range of integration of the $\xi$--fields which can be
found by determine how many times the space of $X^i$'s covers the
$\xi$--space. This may be specified by counting the number of times the
$\xi$-fields pass through zero and in which direction. The zeroes of
the $\xi^i(X)$ are given by the instanton and anti--instanton configurations
\bb
\frac{dX^i}{d\tau}&=&\frac{1}{4!} \{X^{i_1},\cdots,X^{i_5}\}
{\epsilon^i}_{{i_1}\cdots{i_5}}\,\, , \label{inst} \\
 \frac{dX^i}{d\tau}&=&-\frac{1}{4!} \{X^{i_1},\cdots,X^{i_5}\}
{\epsilon^i}_{{i_1}\cdots{i_5}}\, . \label{ant}
\ee
One may easily check that these configurations satisfy the field equations
eq.(\ref{eq}) and that they are absolute minima of the action (\ref{5brane1}).
Fields obeying eq.(\ref{inst}) have $Q\geq 0$  (instantons) while
fields obeying eq.(\ref{ant}) have $Q<0$  (anti--instantons). We expect that
solutions to eqs.(\ref{inst},\ref{ant}) will exist for all Q and thus
the winding number of the Nicolai map is infinity.

\section{The five--brane ground state}

Now, we define the ``superpotential" $W$ through the equation
\bb
\xi^i=\frac{dX^i}{d\tau}\pm\frac{\delta W}{\delta X^i}\, , \label{W}
\ee
which is the most famous of the stochastic equations,
 the Langevin equation. It reflects the
relation of the Nicolai map to the stochastic process of the classical
Euclidean vacuum \cite{st}--\cite{PS}.
Using the Fokker--Planck equation for eq.(\ref{W}) \cite{PW} one can
show that if $W(X)\rightarrow \pm\infty$ as $|X^i|\rightarrow\infty$ then there
exists a large--time limit corresponding to thermal equilibrium.
In this case, the  probability distributions
$P_\pm[X^i,\tau]$, which obey the appropriately regularized
Fokker--Planck equation
\bb
\frac{\partial P_\pm}{\partial\tau}=\int d^5\sigma\frac{\delta}{\delta X^I}
\left(\pm \frac{\delta W}{\delta X^I}+\frac{\delta}{\delta X^I}\right)P_\pm\, ,
\label{distr}
\ee
satisfy
\bb
\lim_{\tau\rightarrow\infty}P_{\pm}[X^i,\tau]=
|\Psi_0^{\pm}(X)|^2
\, , \label{rho}
\ee
where
\bb
\Psi_0^{\pm}(X)&=& C_{\pm}e^{\mp W(X)} \, , \label{Psi} \\
|C_{\pm}|^2&=&\int [dX^i]e^{\mp W(X)}\, , \nonumber
\ee
is the zero--energy supersymmetric ground state, provided that it is
normalizable \cite{st}.
In our case, by solving
\bb
\frac{\partial W}{\partial X^j}=
\frac{1}{5!} \{X^{i_1},\cdots,X^{i_5}\} \epsilon_{j{i_1}\cdots{i_5}}\, ,
\label{WW}
\ee
one may easily verify that the superpotential $W$ is
\bb
W=\frac{1}{6!}\int d^5\sigma X^j
\{X^{i_1},\cdots,X^{i_5}\} \epsilon_{j{i_1}\cdots{i_5}} \, . \label{Wsol}
\ee
Thus,  for the five--brane there exist the
$SO(6)\!\times\!SO(2)$--invariant  vacuum states
\bb
\Psi_0^{\pm}(X)\sim \exp\left[\mp \frac{1}{6!}\int d^5\sigma X^j
\{X^{i_1},\cdots,X^{i_5}\} \epsilon_{j{i_1}\cdots{i_5}}\right]
\, . \label{Psi1}
\ee
These states correspond to   ``forward" and ``backward" stochastic
processes. One of the states (\ref{Psi1})
is in addition normalizable and thus, there
exists zero--energy ground state. It has zero fermion charge and
corresponds to the $N=1$ vector multiplet in ten dimensions \cite{BPS}.

We may also generalize the above
discussion for the other p-branes of the brane scan.
Let us consider a p--brane in D dimensions. In the light--cone gauge there
exist $D\!-\!p\!-\!1$ degrees of freedom describing  transverse excitations.
In this case, there exists the Nicolai map
\bb
\xi^i&=& \frac{dX^i}{d\tau}\pm\frac{1}{p!} \{X^{i_1},\cdots,X^{i_p}\}
{\epsilon^i}_{{i_1}\cdots{i_p}}\, ,~~~(i=1,\cdots,p+1) \label{px}
\ee
analogous to eq.(\ref{x}),
where $\{X^{i_1},\cdots,X^{i_p}\}$ is the  Nambu ``p-bracket"
\bb
\{X^{I_1},\cdots,X^{I_p}\}=\varepsilon^{{a_1}\cdots{a_p}}\partial_{a_1}X^{I_1}
\cdots\partial_{a_p}X^{{I_p}} \, . \label{Nambup}
\ee
The superpotential $W_p$ turns out then to be
\bb
W_p=\frac{1}{(p+1)!}\int d^5\sigma X^j
\{X^{i_1},\cdots,X^{i_p}\} \epsilon_{j{i_1}\cdots{i_p}} \, . \label{Wsolp}
\ee
Then, the   $SO(p+1)\!\times\!SO(D-p-3)$--invariant
ground state is given by eq.(\ref{Psi}) and it is normalizable for $p=3,5$.
The wave--function for $p=2$, in particular, is
\bb
\Psi_0^\pm\sim \exp\left(\mp\frac{1}{6}\int d^2\sigma X^i
\{X^{j},X^{k}\} \epsilon_{ijk}\right)\, ,
\ee
which is non--normalizable and has  been given by de Wit \cite{W}.
Thus, only the three-
and five--brane seem to have a zero--energy ground state.

It should be noted that, in particular for the membrane,
following this method
one may also construct a $G_2$--invariant vacuum wave function.
A Nicolai map for this case may be chosen to be
\bb
\xi^i=\frac{dX^i}{d\tau}\pm \frac{1}{2}
{c^i}_{jk}\{X^j,X^k\}\, , ~~~(i=1,\cdots,7) \, , \label{G2}
\ee
where $c_{ijk}$ are the octonionic structure constants
\cite{GG}. The ``forward"
ground state wave function is then
\bb
\Psi_0^+\sim \exp\left(-\frac{1}{6}\int d^2\sigma X^i
\{X^{j},X^{k}\} c_{ijk}\right)\, .
\ee
It is non--normalizable and coincides with what was reported
in \cite{WHN}. Details will be given elsewhere.

\section{Conclusions}

The purpose of this work has been to report some
results concerning  the ground state of supersymmetric five--branes.
It was initiated by the fact that
although much is known about the ground state
of membranes, similar results for the other branes are lacking.
Based on the ``p-brane democracy" \cite{dem}
as follows from U--duality arguments \cite{U},
the spectrum of all branes are equally important. However,
to determine the spectrum of extended objects other than strings
is a notoriously treacherous subject and one may only
deal with their ground states at the moment.
For the membrane, it seems that there are no massless particles
since there is no zero--energy ground state at least in the quantum mechanical
models considered so far. Claims about the opposite have also been made.

To find the ground state of the five--brane one may follow the standard way of
solving the corresponding Schr\"odinger equation as in the case of the
membrane. Here, however, we chose another, indirect way, in order to find the
vacuum state. Namely, we tried to form a Nicolai map for the theory and then
to read off the vacuum wave function in the equilibrium limit. The existence of
the  latter is equivalent to the existence of normalizable zero--energy ground
state and thus after constructing the Nicolai map, one may check if such a
state exists. We showed this explicitly for the five--brane and we argued
that only the three--brane besides the five--brane has a normalizable
zero--energy ground state. The wave function we found
here is valid up to third order in inverse brane tension and there will
be higher order corrections which however do not spoil its normalizability.

As a final comment, let us note that we studied
here the neutral five--brane since the action of the heterotic one
is not known. Although the former is anomalous \cite{BP},
we expect our results
for the ground state to carried over the heterotic five--brane.

\vspace{1cm}

\noindent
{\large \bf Acknowledgment}
\vspace{.4cm}

\noindent
This work was supported in part by the European Commission TMR  programmes
contract no.
ERBFMRX-CT96-0045 and ERBFMRX-CT96-0090 and the
Greek Secretariat of Research and Technology contract  PENED95/1981.
The author acknowledges the Alexander von Humboldt--Stiftung for financial
support.

\vspace{1cm}

\noindent
{\large\bf Appendix}\setcounter{equation}{0}
\def\theequation{A.\arabic{equation}}
The $SO(8)$ $\gamma$--matrices we use here are
\bb
\Gamma_I=\left(\begin{array}{ll}
0&\gamma_I\\
\gamma_I^\dagger&0 \end{array}\right) \, ~,~~~~I=1,\cdots,8 \, ,
\ee
where  $\gamma_I$ are $8\!\times\!8$ real matrices and
$\gamma_I\gamma_J^\dagger+\gamma_J^\dagger\gamma_I=2\delta_{IJ}$.
One of the $\gamma_I$ can be chosen to be Hermitian
while the rest are anti-Hermitian. In particular
\bb
\gamma_i&=&i\bar{\gamma}_i\, ,~~~ i=1,\cdots,6 \nonumber \\
\gamma_7&=&i\bar{\gamma}_7\,  , \nonumber \\
\gamma_8&=&1 \, ,
\ee
where $\bar{\gamma_i}$ are $SO(6)$ $\gamma$--matrices and
$\bar{\gamma_7}=i\bar{\gamma_1}\cdots\bar{\gamma_6}$ is the
corresponding chiral matrix.

\end{document}